\documentclass[12pt]{article}
\usepackage{epsfig}

\newcommand{\U}{{\cal U}}
\newcommand{\beq}{\begin{equation}}
\newcommand{\eeq}{\end{equation}}
\newcommand{\beqa}{\begin{eqnarray}}
\newcommand{\eeqa}{\end{eqnarray}}

\def\noj#1,#2,{{\bf #1} (19#2)\ }
\def\jou#1,#2,#3,{{\sl #1\/ }{\bf #2} (19#3)\ }
\def\ann#1,#2,{{\sl Ann.\ Physics\/ }{\bf #1} (19#2)\ }
\def\cmp#1,#2,{{\sl Comm.\ Math.\ Phys.\/ }{\bf #1} (19#2)\ }
\def\ma#1,#2,{{\sl Math.\ Ann.\/ }{\bf #1} (19#2)\ }
\def\ng#1,#2,{{\sl Nagoya.\ Math.\ J.\/ }{\bf #1} (19#2)\ }
\def\jd#1,#2,{{\sl J.\ Diff.\ Geom.\/ }{\bf #1} (19#2)\ }
\def\invm#1,#2,{{\sl Invent.\ Math.\/ }{\bf #1} (19#2)\ }
\def\cq#1,#2,{{\sl Class.\ Quantum Grav.\/ }{\bf #1} (19#2)\ }
\def\cqg#1,#2,{{\sl Class.\ Quantum Grav.\/ }{\bf #1} (19#2)\ }
\def\ijmp#1,#2,{{\sl Int.\ J.\ Mod.\ Phys.\/ }{\bf A#1} (19#2)\ }
\def\jmphy#1,#2,{{\sl J.\ Geom.\ Phys.\/ }{\bf #1} (19#2)\ }
\def\jams#1,#2,{{\sl J.\ Amer.\ Math.\ Soc.\/ }{\bf #1} (19#2)\ }
\def\grg#1,#2,{{\sl Gen.\ Rel.\ Grav.\/ }{\bf #1} (19#2)\ }
\def\mpl#1,#2,{{\sl Mod.\ Phys.\ Lett.\/ }{\bf A#1} (19#2)\ }
\def\nc#1,#2,{{\sl Nuovo Cim.\/ }{\bf #1} (19#2)\ }
\def\np#1,#2,{{\sl Nucl.\ Phys.\/ }{\bf B#1} (19#2)\ }
\def\pl#1,#2,{{\sl Phys.\ Lett.\/ }{\bf #1B} (19#2)\ }
\def\pla#1,#2,{{\sl Phys.\ Lett.\/ }{\bf #1A} (19#2)\ }
\def\pr#1,#2,{{\sl Phys.\ Rev.\/ }{\bf #1} (19#2)\ }
\def\prd#1,#2,{{\sl Phys.\ Rev.\/ }{\bf D#1} (19#2)\ }
\def\prl#1,#2,{{\sl Phys.\ Rev.\ Lett.\/ }{\bf #1} (19#2)\ }
\def\prp#1,#2,{{\sl Phys.\ Rept.\/ }{\bf #1C} (19#2)\ }
\def\ptp#1,#2,{{\sl Prog.\ Theor.\ Phys.\/ }{\bf #1} (19#2)\ }
\def\ptpsup#1,#2,{{\sl Prog.\ Theor.\ Phys.\/ Suppl.\/ }{\bf #1} (19#2)\ }
\def\rmp#1,#2,{{\sl Rev.\ Mod.\ Phys.\/ }{\bf #1} (19#2)\ }
\def\yadfiz#1,#2,#3[#4,#5]{{\sl Yad.\ Fiz.\/ }{\bf #1} (19#2) #3%
\ [{\sl Sov.\ J.\ Nucl.\ Phys.\/ }{\bf #4} (19#2) #5]}
\def\zh#1,#2,#3[#4,#5]{{\sl Zh.\ Exp.\ Theor.\ Fiz.\/ }{\bf #1} (19#2) #3%
\ [{\sl Sov.\ Phys.\ JETP\/ }{\bf #4} (19#2) #5]}

\begin{document}
\vspace*{-.6in}
\thispagestyle{empty}
\begin{flushright}
IASSNS-HEP-98/35\\
hep-th/9805029
\end{flushright}
\baselineskip = 20pt

\vspace{.5in}
{\Large
\begin{center}
M-theory realization of a N=1 supersymmetric chiral gauge theory 
in four dimensions\footnote{Work
supported in part by the U.S. Dept. of Energy under Grant No.
DE-FG02-90-ER40542.}
\end{center}}

\begin{center}
Jaemo Park\footnote{E-mail:  jaemo@sns.ias.edu} \\
\emph{School of Natural Sciences, Institute for Advanced Study \\ 
Princeton, NJ 08540 , USA}
\end{center}

\vspace{1in}

\begin{center}
\textbf{Abstract}
\end{center}
\begin{quotation}
\noindent We study M5-brane configuration of the chiral gauge theory 
whose Type IIA brane configuration with orientifold 6-plane(O6) 
is studied by various authors. Much of the paper is devoted to 
studying M-theory picture of SO/Sp gauge theory with fundamental 
flavors realized in Type IIA setup with O6-plane. 
The Higgs branch of N=2 SO/Sp gauge theory is studied and the curve 
corresponding to rotated brane configuration is presented. 
In the chiral gauge theory, the middle NS$^{\prime}$5-brane on top 
of the O6 plane is realized as a detached rational curve. 
Depending on a location of the rational curve in $x^7$ direction,
 the same curve plus 
the rational curve  
can be interpreted as describing the Coulomb branch of 
$SU(2N_c)$ chiral gauge theory, $SO(2N_c)/Sp(N_c)$ gauge theory 
with $N_f/(N_f+4)$ 
fundamental hypermultiplets. Various consistency checks for this 
proposal are made. By introducing two more rational curves corresponding 
to NS$^{\prime}$5-branes, 
one can produce a non-trivial fixed point which mediates chiral 
non-chiral transition. 

\end{quotation}
\vfil

\newpage

\pagenumbering{arabic}

\section{Introduction}

Recently there has been much progress in understanding non-perturbative 
aspects of supersymmetric gauge theories using brane configuration. 
The configuration where D-branes are suspended between NS-branes 
was firstly studied in \cite{HW}, and the subsequent study of similar 
configurations in various dimensions deepen our understanding on 
supersymmetric gauge theories \cite{Gi}.
  
Especially interesting examples are the brane realizations of 
supersymmetric gauge theories in 
four dimensions with four or eight supercharges. 
One particular success 
is that the N=1 Seiberg's duality 
can be reproduced in this approach. 
Much of the gauge theories studied via brane setup are non-chiral, 
but there have been fruitful attempts to understand chiral 
gauge theories using brane dynamics\cite{Hanany2, Trivedi, Hanany4}. 
For four dimensional gauge theories, one can lift the brane setup in 
Type IIA to the eleven dimensions where D4-branes and NS-5 branes 
altogether turn into M-5 branes\cite{Witten1}. 
Several aspects of supersymmetric 
gauge theories are clear in the eleven dimensions such as the Riemann 
surface associated with the Coulomb branch of a specific gauge theory. 

Even though there are several constructions of chiral gauge theories 
in ten-dimensional brane configurations, so far the corresponding 
eleven-dimensional construction is lacking. 
The purpose of this paper is to propose the M-theory realization 
of a particular chiral gauge theory constructed in 
\cite {LLL, Ha3, Kuta}. 
Orientifold 6-plane is introduced to construct the chiral gauge theory. 
The brane configuration with orientifold plane leads to the 
understanding of the SO/Sp gauge theories as well as the construction 
of SU gauge theories with symmetric/antisymmetric matter\cite{Johnson,LL}.
One can introduce either orientifold 4-plane or orientifold 6-plane
(O6) 
in the brane setup without breaking further supersymmetry, but the 
relevant one for our study is orientifold 6-plane.  
Much of the paper will be devoted to understanding the SO/Sp 
gauge theories with orientifold 6-plane, since this is closely related 
to the study of the chiral gauge theory, as we will see. 
 
The content of this paper is as follows. 
In section 2, we introduce the basic brane setup with orientifold 
6-plane and review how to construct the chiral gauge theory. 
In section 3, we study M-theory interpretation of SO/Sp gauge theories 
with fundamental flavors and see 
how the Higgs branch of the N=2 gauge theories 
are realized in M-theory setup. This constitutes a part of the 
moduli space of the chiral gauge theory. 
In section 4, we propose the rotated N=1 brane configuration 
of SO/Sp gauge theories starting from N=2
brane setup. 
In section 5, we propose the 11-d interpretation of the chiral 
gauge theory. Closely related chiral-nonchiral transitions are discussed. 

After we obtained the results in section 4, we receive the preprint 
where the same problem is solved by a slightly different approach
\cite{Ahn2}. 
We understand that the M5-brane construction  
 of the chiral gauge theory treated in this paper is known to 
\cite{LLL2}.

\medskip  
 \section{Relevant Brane Configuration with Orientfold Plane}

 In order to have N=2 supersymmetry in 4-d, we need three ingredients 
 in Type IIA string theory; NS5-brane whose world volume spans 
$x^0,x^1,x^2,x^3,x^4,x^5$, D4-branes whose world volume spans
  $x^0,x^1,x^2,x^3,x^6$, and D6-branes whose world volume spans 
$x^0,x^1,x^2,x^3,x^7,x^8,x^9$. The $SO(1,9)$ Lorentz symmetry is broken 
to $SO(1,3)\times SO(2)_{4,5} \times SO(3)_{7,8,9}$ 
where $SO(n)_{i_1, \ldots i_n}$ 
is a rotational group in $x^{i_1}, \ldots x^{i_n}$-direction.  
$SO(3)_{7,8,9}$ 
can be identified with $SU(2)_R$ 
symmetry of the N=2 gauge theory. We can introduce O6-plane parallel 
to D6-brane without breaking supersymmetry further. We have two types 
of O6-plane, one carries +4 Ramond-Ramond(RR) charge and the other 
carries --4 RR charge if we normalize the RR charge of a D4-brane 
to be +1. 
The former gives SO gauge group on the world volume of D4-brane and the 
latter gives Sp gauge group. In the presence of O6-plane, 
the configuration should be symmetric under the $Z_2$ action 
$(x^4,x^5,x^6)\rightarrow (-x^4,-x^5,-x^6)$. Throughout the paper, we 
use the covering space of the $Z_2$ action when we count the number 
of branes. 
Thus if we have $2N_c$ D4-branes and $2N_f$ D6-branes, 
the resulting gauge group 
is $SO(2N_c)/Sp(N_c)$ with $N_f$ N=2 hypermultiplets 
where $Sp(N_c)$ has rank $N_c$. 

One can rotate a NS5-brane by an arbitrary angle $\theta$, which 
characterizes the rotation between $(x^4,x^5)$-plane and 
$(x^8,x^9)$-plane. 
The supersymmetry is reduced to N=1. The rotated 5-brane with 
$\theta=\pi/2$ is denoted by NS$^{\prime}$5-brane whose world volume 
is in the $x^0,x^1,x^2,x^3,x^8,x^9$-direction. $SO(1,9)$ spacetime 
Lorentz symmetry is broken to 
$SO(1,3)\times SO(2)_{4,5}\times SO(2)_{8,9}$. The first factor is the 
Lorentz symmetry of the 4-d theory. The two $SO(2)$ factors can be 
identified as $R$ symmetries of the N=1 theory. 
\begin{figure}
\centering
\epsfxsize=4in
\hspace*{0in}
\epsffile{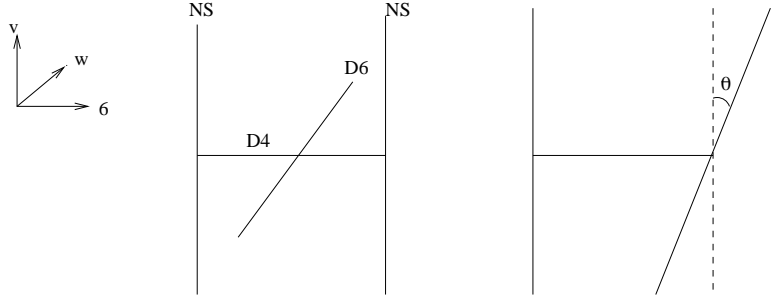}
\caption{The left figure illustrates the basic brane configuration 
with N=2 supersymmetry. $v=x^4+ix^5$ and $w=x^8+ix^9$. 
In the right figure 
the right NS5-brane is rotated by $\theta$.}
\label{fig:brane}
\end{figure}

\begin{figure}
\centering
\epsfxsize=5in
\hspace*{0in}
\epsffile{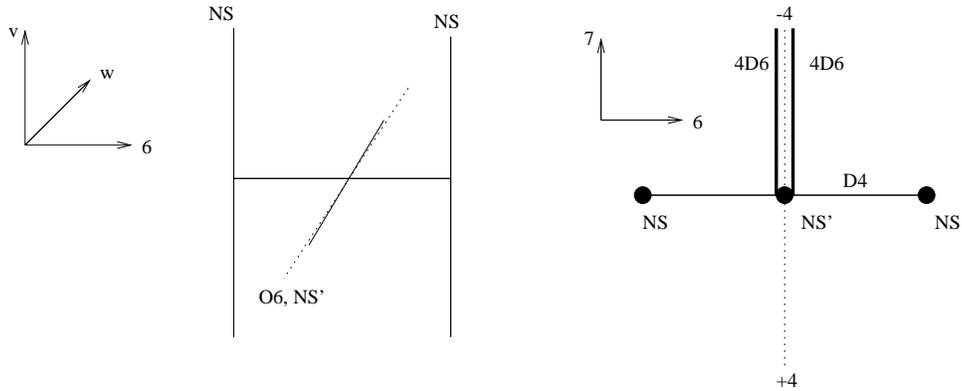}
\caption{The configuration for the chiral gauge theory. Left figure 
describes the brane configuration in $(v,w,x^6)$-plane while right 
figure describes same configuration in $(x^6,x^7)$-plane. Note that 8
half D6-branes are stuck on the half of O6-plane.}
\label{fig:brane2}
\end{figure}

The chiral gauge theory we are interested in is constructed by placing 
NS$^{\prime}$5-brane on top of the O6-plane. O6-plane changes its RR 
charge as it traverses the NS$^{\prime}$5-brane. This is the T-dual 
version of \cite{Johnson} where it is shown that O4-plane changes 
its RR charge as it traverses a NS5-brane. Anomaly cancellation of the 
corresponding field theory or the charge conservation requires that 
4 physical half D6-branes and the mirror are stuck on the negatively 
charged half of the O6-plane. If we have $2N_c$ D4-branes, the gauge 
group is $SU(2N_c)$ since the orientifold identifies the left 
$SU(2N_c)_L$ and the right $SU(2N_c)_R$. Among the degrees of freedom 
that lead to the bifundamentals in the product gauge group
$SU(2N_c)_L\times SU(2N_c)_R$, only one 
antisymmetric and one conjugate symmetric tensor survive under the 
orientifold projection.\footnote{We enumerate bosonic degrees of freedom
in each N=1 supermultiplet. }
Thus we have $SU(2N_c)$ gauge group with one 
antisymmetric tensor $A$, one conjugate symmetric tensor $\tilde{S}$ and 
8 fundamentals $T$ which come from 8 half D6-branes stuck on the half of 
O6-plane. This is indeed an anomaly free chiral theory. In the presence 
of D6-branes we have vectorlike fundamental flavors $Q, \tilde{Q}$. 

We can rotate left NS5-brane by $\theta$ and right NS5-brane 
by $-\theta$. The superpotential 
is given by 
\begin{equation}
W=\tilde{S}TT+mX^2+XA\tilde{S}, \label{eq:pot}
\end{equation}
where $m=tan\theta$ and $X$ is the adjoint of $SU(2N_c)$. 
The first term can be deduced by considering the flat directions as we 
will see shortly. \footnote{Alternatively one can consider 
the global symmetry. The flavor symmetry coming from 8 half D6-branes 
is $SO(8)$ and the first term of (\ref{eq:pot}) should be introduced 
to have the right flavor symmetry.} 
The remaining terms are the superpotential of the product gauge group 
which are invariant under the orientifold projection. 
The resulting superpotential after integrating out the massive adjoint 
in case $m \neq 0$ is 
\begin{equation}
W=\tilde{S}TT-\frac{1}{2m}(A\tilde{S})^2.
\end{equation}
One can easily incorporate the superpotential due to the presence 
of D6-branes. 

In the limit $m\rightarrow \infty$, we have Coulomb branch and $SU(2N_c)$ 
is generically broken to $U(1)^{N_c}$. In the brane picture, 
this corresponds to pairwise movement of D4-branes in 
$x^4,x^5$-direction compatible with 
the orientifold projection. In the Coulomb branch, the D4-branes 
are reconnected and stretched between the NS5-brane and its mirror. 
If $m$ is not zero, the mass of the adjoint 
does not allow any Coulomb branch. But we have several Higgs branches. 
Meson branches arise in the usual way by splitting D4-branes between 
D6-branes. In addition to that, there are interesting baryonic branches. 
In the flat directions where the baryon operator 
$A^n\tilde{Q}^{2N_c-2n}$ gets an expectation value, the gauge group 
is broken to $Sp(n)$ with a symmetric tensor(adjoint) with a 
superpotential $W=mX^2$. In the brane picture, this corresponds to 
moving NS$^{\prime}5$-brane in the positive $x^7$-direction. 
On the other hand, in the flat direction where the 
baryon operator $\tilde{S}^n Q^{2N_c-n}$ gets an expectation value, 
the gauge group is reduced to $SO(2n)$ with an adjoint with $W=mX^2$. 
This corresponds to moving NS$^{\prime}$5-brane in the negative 
$x^7$-direction. The fundamentals arising from half D6-branes become 
massive, as is obvious from the brane picture. In case $m=0$, these 
baryonic branches have an accidental N=2 supersymmetry.

\medskip
\section{Higgs branch of N=2 SO/Sp gauge theories}
The M-theory lifting of the brane configuration with O6-plane 
was discussed in \cite{LL}. O6-plane with $-4 RR$ charge ($O6^-$) is 
lifted to Atiyah-Hitchin space in M-theory\cite{SeWi, Se}. 
For O6-plane with +4 RR charge ($O6^+$), the corresponding M-theory 
interpretation is less clear, but it appears to be a $D_4$ singularity 
which cannot be blownup \cite{Witten3}. In order to describe the 
curve for the Coulomb branch of a gauge theory 
we just need to concentrate 
on the complex structure of these spaces. For $O6^-$ plane, far away from 
the origin, it can be described by 
\begin{equation}
xy=\Lambda^{4N_c+4}_{N=2}v^{-4}.   \label{eq:O61}
\end{equation}
In the usual convention of M-theory setting $v=x^4+ix^5, 
t=exp-\frac{x^6+ix^{10}}{R}$ where $R$ is the radius of the 11-th circle
, we have $x\sim t^{-1}$ for a finite value of $y$ and $y\sim t$ for 
a finite value of $x$.   

The curve for pure $Sp(N_c)$ gauge theory can be written as 
\begin{equation}
x+y=B_{N_c}(v^2)+\Lambda^{2N_c+2}_{N=2}
\frac{2}{v^2}, 
\end{equation}
or 
\begin{equation}
y^2-y(B_{N_c}(v^2)+\Lambda^{2N_c+2}_{N=2}\frac{2}{v^2})+
\Lambda^{4N_c+4}_{N=2}v^{-4}=0
\end{equation}
using (\ref{eq:O61}). Here $B_{N_c}(x)$ is a polynomial of degree 
$N_c$ in $x$ and $\frac{1}{v^2}$ term is allowed due 
to $v^{-4}$ contribution 
of the orientifold. We can interpret $\frac{1}{v^2}$ term as a 
manifestation that close to $v=0$, (\ref{eq:O61}) does not provide a 
good description of the orientifold background\cite{LL}. 
We can incorporate 
the $2N_f$ D6-branes by considering 
\begin{equation}
xy=\Lambda^{4N_c+4-2N_f}_{N=2}v^{-4}
\prod_{i=1}^{f}(v^2-m_i^2). \label{eq:D61}
\end{equation}
Then the curve for the $Sp(N_c)$ gauge theory with N=2 $N_f$ 
hypermultiplet is given by
\begin{equation}
x+y=B_{N_c}(v^2)+\Lambda^{2N_c+2-N_f}_{N=2}
\frac{2\prod_{j=1}^{f}im_j}{v^2} \label{eq:sp}
\end{equation}
or 
\begin{equation}
y^2-y(B_{N_c}(v^2)+\Lambda^{2N_c+2-N_f}_{N=2}
\frac{2\prod_{j=1}^{f}im_j}{v^2})
+\Lambda^{4N_c+4-2N_f}_{N=2}v^{-4}\prod_{j=1}^{f}(v^2-m_j^2)=0.
\label{eq:Sp}
\end{equation}
Note that (\ref{eq:sp}) is invariant under the $Z_2$ projection 
$v\rightarrow -v, x\leftrightarrow y$. Again nonzero $\frac{1}{v^2}$ 
term tell us that (\ref{eq:O61}) is not a good description near $v=0$. 

The Type IIA brane configuration is invariant under the rotations 
$SO(2)_{4,5}$ and $SO(3)_{7,8,9}$ if the order parameters are 
also rotated accordingly. These can be interpreted as classical 
$U(1)$ and $SU(2) R$-symmetry of the 4-d world volume theory of the 
brane. In the M-theory construction, $SU(2)_{7,8,9}$ is preserved 
but $U(1)_{4,5}$ is broken. We can preserve the discrete subgroup 
$Z_{4N_c+4-2N_f}$ of $U(1)_{4,5}$ if we modify the $U(1)_{4,5}$ action 
so that the variables $x$ and $y$ have charges $4N_c$. The full $U(1)$ 
symmetry is restored if we assign the instanton charge $(4N_c+4-2N_f)$ 
to the factor $\Lambda^{2N_c+2-N_f}_{N=2}$, reflecting the $U(1)_R$ 
anomaly. The modified $U(1)_{4,5}$ charges are 
\begin{equation}
\begin{array}{cccc}
x & y & v & \Lambda^{2N_c+2-N_f}_{N=2} \\
4N_c & 4N_c & 2 & 4N_c+4-2N_f 
\end{array}
\end{equation}

For N=2 $Sp(N_c)$ gauge theory, the flavor symmetry is $SO(2N_f)$. 
But from (\ref{eq:D61}), $SO(2N_f)$ symmetry is not clear. 
This can be seen by some variable changes. 
In case all $m_i$ vanish in (\ref{eq:D61}), we have 
\begin{equation}
xy=v^{-4+2N_f} \label{eq:Dnf}
\end{equation}
if we set $\Lambda_{N=2}=1$ for simplicity. 
If we perform the variable change 
\begin{equation}
\tilde{x}^{\prime}=\frac{-x+y}{2}v, \tilde{y}^{\prime}=-(x+y), 
\label{eq:new} 
\end{equation} 
(\ref{eq:Dnf}) becomes 
\begin{equation}
(\tilde{x}^{\prime})^2=\frac{v^2}{4}(\tilde{y}^{\prime})^2
-v^{-2+2N_f}.  \label{eq:Dnf2}
\end{equation}
This can be recast to the standard $D_{N_f}$ singularity 
\begin{equation}
x^2+y^2z=z^{n-1}
\end{equation}
by setting $z=v^2, x=i\tilde{x}^{\prime}$ and 
$y=\frac{\tilde{y}^{\prime}}{2}$. 
In case $N_f=0$, (\ref{eq:Dnf2}) becomes 
$(\tilde{x}^{\prime})^2=\frac{z}{4}(\tilde{y}^{\prime})^2-\frac{1}{z}$. 
With the variable change $\tilde{y}^{\prime}=\tilde{y}+
\frac{2}{z}$, we have
\begin{equation}
 (\tilde{x}^{\prime})^2=\frac{z}{4}(\tilde{y}^{\prime})^2+y 
\label{eq:AH}.
\end{equation}
(\ref{eq:AH}) defines the Atiyah-Hitchin space as a complex manifold 
in one of its complex structure\cite{AH}.  
With the new variables (\ref{eq:new}), the curve for the $Sp(N_c)$ 
gauge theory can be written as 
\begin{equation}
\tilde{y}^{\prime}=B_{N_c}(z)  \label{eq:S}
\end{equation}

If all $m_i$ in (\ref{eq:D61}) vanish, corresponding D6-branes are 
located at the same position in $x^4,x^5$-direction, but they can be 
separated in the $x^6$-direction. When all $m_i$ vanish, we saw that 
the surface develops $D_{N_f}$-singularity. The separation of the 
D6-brane in the $x^6$-direction corresponds to the resolution of 
singularity. The resolution of $D_{N_f}$ singularity makes it 
possible to identify the Higgs branch of N=2 $Sp(N_c)$ gauge theory with 
$N_f$ flavors. This will be the topic of the next subsection.

Note that the variable change 
(\ref{eq:new}) is singular at $x=y=v=0$. 
Actually the description $xy=v^{-4+2N_f}$ is not valid near the origin 
and we have some difficulty in understanding the blowup just from this 
approximate description. 
As above, if we go to the correct description, we can straightforwardly 
obtain the resolution of $D_{N_f}$. One interesting work is done in 
\cite{Sen} in light of this subtlety. The Atiyah-Hitchin space 
looks like a Taub-Nut space with a negative mass parameter 
asymptotically. 
The M-theory configuration corresponding $O6^-$-plane with D6-branes 
can be described approximately by a Taub-Nut space with a negative 
mass parameter superposed by multi Taub-Nut space corresponding to 
D6-branes. There are several two cycles in this space and 
the intersection matrix of the independent cycles is shown to be 
that of $D$-singularities. In one of the complex structures, this 
space can be described by the equation $xy=v^{-4+2N_f}$ if the 
number of D6-branes is $2N_f$.

For $SO(2N_c)$ gauge theory, the corresponding $O6^+$ plane is 
described by $xy=\Lambda_{N=2}^{4N-c-4}v^4$ 
and in the presence of $2N_f$ D6-branes 
the configuration is described by 
\begin{equation}
xy=\Lambda^{4N_c-4}_{N=2}
v^4\prod_{i=1}^{N_f} (v^2-m_i^2) \label{eq:xyso}
\end{equation} 
in M-theory. The Coulomb branch of 
$SO(2N_c)$ gauge theory with $N_f$ N=2 hypermultiplets is described 
by the curve 
\begin{equation}
x+y=B_{N_c}(v^2), \label{eq:so}
\end{equation}  
or
\begin{equation}
y^2-B_{N_c}(v^2)y+\Lambda^{4N_c-4+2N_f}_{N=2}
v^4\prod_{i=1}^f (v^2-m_i^2)=0.  \label{eq:so2}
\end{equation}
\subsection{Resolution of $D_n$ singularity}

The (mixed Coulomb-) Higgs branches of the $Sp(N_c)$ gauge theory 
with $N_f$ fundamentals were analyzed in \cite{Argyres}. 
They are classified by an integer $r=1,2 \ldots [\frac{N_f}{2}]$. 
The $r$-th Higgs branch has quaternionic dimension $2rN_f-(2r^2+r)$ 
and emanates from a $(N_c-r)$-dimensional complex subspace of the 
Coulomb branch where there is an unbroken gauge group $Sp(r)$. 
Higgs branches are not corrected by quantum effects, but there are 
interesting subtleties in the way the Higgs branches emanates from 
the Coulomb branch. In the classical picture, Higgs branch emanates 
from the locus where $r$ of the $\phi_a$'s vanish where $\phi_s$'s 
appearing in $B_{N_c}(v^2)=\prod_{a=1}^{N_c}(v^2-\phi_a^2)$. 
This will be true quantum mechanically for values $r$ where $Sp(r)$ 
gauge theory with $N_f$ flavors is not asymptotically free, i.e., 
for $r \leq \frac{N_f-2}{2}$. For $r=[\frac{N_f}{2}]$, the low 
energy theory is asymptotically free and can be affected by strong 
coupling effects. We will see that the Higgs branch emanates from the 
locus where only $\frac{N_f-2}{2}$ of the $\phi_a$'s vanish and the 
product of non-zero $\phi_a^2$ is $\pm 2\Lambda^{2N_c+2-N_f}$. 
We will see that the above field theory results are consistent 
with M5-brane picture. In order to do show that, we should understand 
the resolution of $D_{N_f}$-singularity.  

The blowup of $D_{N_f}$ surface is explained in detail at \cite{Hori}, 
and we just quote their results. Starting from the singular surface 
of type $D_{N_f}, 
x^2+y^2 z=z^{N_f-1}$ in $x$-$y$-$z$ space, we obtain a 3-fold covered by 
$N_f$ open subsets $\U_1, \U_2, \cdots \U_{N_f}$ with coordinates  
$(s_1,t_1,z_1)=(x,y,\widetilde{z})$,
$(s_2=y,t_2=\widetilde{x},z_2),$ $\ldots,$ $(s_{N_f},t_{N_f},z_{N_f})$.
These open sets are glued together
by transition relations;
$(s_j,t_j,z_j)=(s_{j+1}t_{j+1}z_{j+1},s_{j+1},t_{j+1}^{-1})$
for $j=1,\ldots,N_f-4$,
$(s_{N_f-3},t_{N_f-3},z_{N_f-3})
=(s_{N_f-2}t_{N_f-2}^2z_{N_f-2},s_{N_f-2}t_{N_f-2},t_{N_f-2}^{-1})$,
and
$(s_{N_f-2},t_{N_f-2},z_{N_f-2})
=(z_{N_f-1}t_{N_f-1},s_{N_f-1},t_{N_f-1}^{-1})
=(t_{N_f}^{-1},s_{N_f}t_{N_f},z_{N_f})$.
The projection to
the $x$-$y$-$z$ space is given by
\begin{equation}
\left\{
\begin{array}{lclcl}
x&=&s_{2j-1}^jz_{2j-1}^{j-1}&=&
s_{2j}^jt_{2j}z_{2j}^j\\[0.2cm]
y&=&s_{2j-1}^{j-1}t_{2j-1}z_{2j-1}^{j-1}
&=&s_{2j}^jz_{2j}^{j-1}\\[0.2cm]
z&=&s_{2j-1}z_{2j-1}&=&s_{2j}z_{2j}
\end{array}  \right.
\label{projD}
\end{equation}
on $\U_1,\ldots,\U_{N_f-3},\U_{N_f-1}$. 
We need the expression of $y,z$ on $\U_{N_f-2}$ and $U_{N_f}$: 
\beqa
&&y=\left\{
\begin{array}{ll}
z^{\frac{N_f}{2}-2}s_{N_f-2}t_{N_f-2}=s_{N_f}z^{\frac{N_f}{2}-2}&
\quad \mbox{$N_f$ : even}\\[0.2cm]
t_{N_f-2}z^{[\frac{N_f}{2}]-1}=s_{N_f}t_{N_f}z^{[\frac{N_f}{2}]-1}&
\quad\mbox{$N_f$ : odd}
\end{array}
\right.
\label{projDy}\\
&&z=s_{N_f-2}t_{N_f-2}z_{N_f-2}=s_{N_f}z_{N_f}.
\label{projDz}
\eeqa

The resolved $D_{N_f}$ surface is given by
\beqa
&&s_i+t_i^2z_i=s_i^{N_f-1-i}z_i^{N_f-i} \qquad
\mbox{in~ $\U_i$}\qquad (i\ne N_f-2,N_f)
\label{Deq1}\\
&&s_{N_f-2}+t_{N_f-2}z_{N_f-2}=s_{N_f-2}z_{N_f-2}^2 
\qquad \mbox{in~ $\U_{N_f-2}$}\\
&{\rm and}&
\quad 1+s_{N_f}t_{N_f}^2z_{N_f}=z_{N_f}^2 
\qquad\qquad \mbox{in~ $\U_{N_f}$}
\label{Deq3}
\eeqa
This is mapped onto the singular $D_{N_f}$ surface by (\ref{projD}), and
the map is an isomorphism except at the inverse image of the
singular point $x=y=z=0$. The inverse image consists of
$N_f$ rational curves $C_1,\ldots,C_{N_f}$ where $C_i$ 
($i=1,\ldots,N_f-2$)
is the $z_i$-axis in $\U_i$ (i.e.,~$s_i=t_i=0$), and also the
$t_{i+1}$-axis in $\U_{i+1}$ (i.e.~$s_{i+1}=z_{i+1}=0$).
$C_{N_f-1}$ and $C_{N_f}$ are the 
loci $t_{N_f-2}=z_{N_f-2}\mp 1=0$ parallel to
the $s_{N_f-2}$-axis in $\U_{N_f-2}$. $C_{i-1}$ and $C_i$
($i=2,\ldots,N_f-2$)
intersects transversely at $s_i=t_i=z_i=0$, while $C_{N_f-2}$
intersects also with $C_{N_f-1}$ and $C_{N_f}$ at
$s_{N_f-2}=t_{N_f-2}=z_{N_f-2}\mp 1=0$.
There is no other intersection of distinct $C_i$'s.
The self-intersection of $C_i$
in the resolved surface can be shown to be $-2$.

\subsection{N=2 Higgs branch of $Sp(N_c)$ gauge theory}
In the Type IIA brane setup, Higgs branch is described by D4-branes 
suspended between D6-branes where they can move in $x^7, x^8, x^9$ 
directions. Likewise, in M-theory the transition to the Higgs branch 
occurs where the 5-brane intersects with the D6-branes. This is 
possible only when 
$y=\frac{1}{2}\tilde{y}^{\prime}=\frac{1}{2}B_{N_c}(z)$ of (\ref{eq:S}) 
passes through the singular point 
$x$=$y$=$z$=$0$. Thus we need $B_{N_c}(z=0)=0$ or 
\begin{equation}
B_{N_c}(z)=z^r(z^{N_c-r}+u_1 z^{N_c-r-1}+\cdots +u_{N_c-r}), \label{eq:r} 
\end{equation}
and we can assume that $u_{N_c-r}$ is non-zero.\footnote{For simplicity,
we set $\Lambda_{N=2}$ to 1 in this subsection otherwise stated.} 
The terms in the bracket describes the Coulomb branch which is broken 
to $U(1)^{N_c-r}$ and this branch dose not meet the D6-brane. 
The term $B_{N_c}(z)=z^r$ near $z=0$ 
describes Higgs branch and we should use 
the resolved $D_{N_f}$ surface in order to describe this. Since higher 
terms of (\ref{eq:r}) are negligible, we can just consider the 
equation $B_{N_c}=z^r$.\footnote{We will follow 
the presentation of \cite{Hori} closely.}  
Let us look at the $F\equiv y-z^r$ in the $2j$-th patch $\U_{2j}, 
2j\leq N_{f}-3$. 
\begin{eqnarray}
F&=& y-z^r \\
 &=& s_{2j}^jz_{2j}^{j-1} (1-s_{2j}^{r-j}z_{2j}^{r-j+1}) \,\,\,  
j\leq r \label{eq:n1}\\
 &=& s_{2j}^r z_{2j}^{r-1}(1-z_{2j})  \,\,\, j=r \label{eq:n2} \\
 &=& s_{2j}^rz_{2j}^r(-1+s_{2j}^{j-r}z_{2j}^{j-1-r}). \,\,\, 
j\geq r+1  \label{eq:n3}\\
\end{eqnarray}   
The curve consists of several components. One component, 
to be denoted by $C$, is the zero of the last factor of the above 
equations (\ref{eq:n1}),(\ref{eq:n2}) and (\ref{eq:n3}). This extends 
to the region away from $x=y=z=0$ reaching infinity.   
$F$ also has zero at $s_{2j}=0$ and $z_{2j}=0$. 
The defining equation of the surface is 
\begin{equation}
s_{2j}+t_{2j}^2z_{2j}=s_{2j}^{N_f-1-2j}z_{2j}^{N_f-2j}.
\end{equation}
There are two branches of zero of $F$ for each $j$; $s_{2j}=z_{2j}=0$ 
and $s_{2j}=t_{2j}=0$ which corresponds to the rational curves 
$C_{2j-1}$ and $C_{2j}$ respectively. 
Near the first branch $s_{2j}=z_{2j}=0$,
$(t_{2j},z_{2j})$ is a good coordinate, i.e.
$s_{2j}$ can be uniquely expressed in terms of $t_{2j}$ and
$z_{2j}$ by the defining equation.
Since $t_{2j}\ne 0$
generically, $s_{2j}\sim z_{2j}$ near $C_{2j-1}$.
Hence
$F\sim z_{2j}^j z_{2j}^{j-1}=z_{2j}^{2j-1}$ for $j\leq r$
while $F\sim z_{2j}^{2k}$ for $j>r$.
Thus, the zero of $F$ at $C_{2j-1}$ is of order $2j-1$ for
$j\leq r$ and order $2r$ for $j>r$.
Near the second branch $s_{2j}=t_{2j}=0$,
$(t_{2j},z_{2j})$ is again a good coordinate, and
$s_{2j}\sim t_{2j}^2$ for $z_{2j}\ne 0$.
Thus, $F\sim t_{2j}^{2j}$ for $j\leq r$
and $F\sim t_{2j}^{2r}$ for $j>r$ near $C_{2j}$.
Namely, $F$
has a zero at $C_{2j}$ of order $2j$ for $j\leq r$
and $2r$ for $j>r$.
By looking at the equation for $j=r$, we see that the infinite curve
$C$ and the rational curve $C_{2r}$ meet at the point
$s_{2r}=t_{2r}=0, z_{2r}=1$.
For ${\cal U}_{2j-1}$ the analysis is similar and we obtain the same 
result.  
In summary,
in the patches
${\cal U}_1,\ldots,{\cal U}_{N_f-3}$,
$F$ has zeros at $C_1,C_2,\ldots,C_{2r-1},C_{2r},C_{2r+1},\ldots,
C_{N_f-3}$ and $C$ of order $1,2,\ldots,2r-1, 2r,2r,\ldots, 2r$ and $1$
respectively.

\begin{figure}
\centering
\epsfxsize=5in
\hspace*{0in}
\epsffile{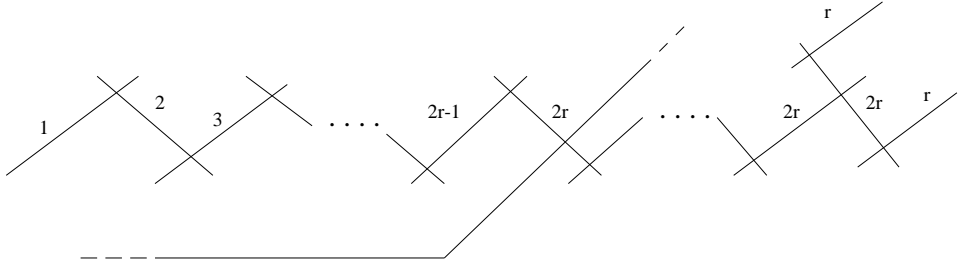}
\caption{The r-th Higgs branch $Sp(N_c)$ gauge theory in M-theory 
with $2r \leq N_f-3$. 
The short lines  
represent the rational curves arising from the resolution of 
$D_{N_f}$ singularity. The long line is the curve $C$ which extends to 
infinity}
\label{fig:dnf}
\end{figure}

Let us now look at the function $F$ in ${\cal U}_{N_f-2}$.
If $2r \leq N_f-3$, from (\ref{projDy})-(\ref{projDz})
we can see that $y$ is divisible by $z^r$,
and $y/z^r-1$ of $F=z^r(y/z^r-1)$ has a single zero at $C$.
Now, we consider the zero of
$z^r=(s_{N_f-2}t_{N_f-2}z_{N_f-2})^r$.
By looking at the defining equation of the surface
$$
s_{N_f-2}(z_{N_f-2}^2-1)=t_{N_f-2}z_{N_f-2}
$$
we see that there are four branches of zero:
$s_{N_f-2}=z_{N_f-2}=0$, $s_{N_f-2}=t_{N_f-2}=0$,
$t_{N_f-2}=z_{N_f-2}-1=0$ and $t_{N_f-2}=z_{N_f-2}+1=0$,
which corresponds to $C_{N_f-3}$, $C_{N_f-2}$, $C_{N_f-1}$ and
$C_{N_f}$ respectively.
Near $C_{N_f-3}$ where $s_{N_f-2}=z_{N_f-2}=0$ and $t_{N_f-2}\ne 0$,
the surface is coordinatized
by $(t_{N_f-2},z_{N_f-2})$
and $F\sim s_{N_f-2}^rz_{N_f-2}^r\sim z_{N_f-2}^{2r}$
has zero at $C_{N_f-3}$ of order $2r$, as we have seen.
Near $C_{N_f-2}$ where
$s_{N_f-2}=t_{N_f-2}=0$ and $z_{N_f-2}\ne 0$,
the surface is again coordinatized
by $(t_{N_f-2},z_{N_f-2})$
and $F\sim s_{N_f-2}^rt_{N_f-2}^r\sim t_{N_f-2}^{2r}$
has zero at $C_{N_f-2}$ of order $2r$.
Near $C_{N_f-1}$ or $C_{N_f}$ where
$t_{N_f-2}=z_{N_f-2}\mp 1=0$ and $s_{N_f-2}\ne 0$,
the surface is coordinatized
by $(s_{N_f-2},z_{N_f-2})$, and
$F\sim t_{N_f-2}^r\sim (z_{N_f-2}\mp 1)^r$ has zero
at $C_{N_f-1}$ and $C_{N_f}$ of order $r$.
The zero of $F$ in the part in ${\cal U}_{N_f-1}$ and ${\cal U}_{N_f}$
can be seen in the same way, and the analysis agrees with that in 
$\U_{N_f-2}$. 
Once the curve degenerates and rational curves are generated, they 
can move in $(x^7, x^8, x^9)$-directions. This motion 
together with the integration of the chiral two-form field  
over such rational curves parametrize the Higgs branch of the 4-d theory
\cite{Witten1}. Since the total rational components are $C_i$ times 
the order of $F$ in $C_i$, 
the quaternionic dimension of the $r$-th Higgs branch 
is $\Sigma_{i=1}^{2r}i+2r\times(N_f-2r-2)
+2r=2rN_f-r(2r+1)$ as expected. 

For $N_f=2r, 2r+1, 2r+2,$ the analysis is a bit different 
since the counting in $\U_{N_f-2}$ is modified. 
For $N_f=2r+2$, with $y=\frac{a}{2}z^r$ the counting in $\U_1, \U_2 
\ldots \U_{N_f-3}$ is the same as before and we have zeroes of order 
$1,2,\ldots k, \ldots N_f-3$ in $C_1,C_2, \ldots C_k, \dots C_{N_f-3}$
respectively.\footnote{Previously 
the coefficient of $z^r$ is not important 
but for $N_f=2r+2$ this is going to be important. The $\frac{1}{2}$
factor comes from (\ref{eq:S}) the relation between $y$ and 
$\tilde{y}^{\prime}, y=\frac{\tilde{y}^{\prime}}{2}$.} 
In $\U_{N_f-2}$, we have 
\begin{equation}
y-\frac{a}{2}z^r=(s_{N_f-2}t_{N_f-2}z_{N_f-2})^{r-1}
(s_{N_f-2}t_{N_f-2}-\frac{a}{2}s_{N_f-2}t_{N_f-2}z_{N_f-2}).
\end{equation}
At $C_{N_f-3}, y-\frac{a}{2}z^r\sim (s_{N_f-2}z_{N_f-2})^{r-1}s_{N_f-2}
\sim z_{N_f-2}^{2r-1}$ which gives zeroes of order ${N_f-3}$ as obtained 
above. 
At $C_{N_f-2}, y-\frac{a}{2}z^r\sim (s_{N_f-2}t_{N_f-2})^r 
(1-\frac{a}{2}z_{N_f-2})
\sim (t_{N_f-2})^{2r}$ and we see that the infinite curve $C$ meet with 
$C_{N_f-2}$ at $z_{N_f-2}=\frac{2}{a}$. 
At $C_{N_f-1}$ where $t_{N_f-2}=z_{N_f-2}-1=0$, 
$y-\frac{a}{2}z^r\sim (t_{N_f-2})^r(1-\frac{a}{2}z_{N_f-2})$ 
and the order of zero is $r$ 
if $a\neq 2$ and $r+1$ if $a=2$. 
At $C_{N_f}$ where $t_{N_f-2}=z_{N_f-2}+1=0$, 
$y-\frac{a}{2}z^r\sim (t_{N_f-2})^r(1-\frac{a}{2}z_{N_f-2})$ 
and the order of zero is $r$ 
if $a\neq -2$ and $r+1$ if $a=-2$.
Thus the quaternionic dimension we get is 
$\frac{N_c(N_c-1)}{2}-1$ if $a \neq \pm 2\Lambda^{2N_c+2-N_f}_{N=2}$
 and 
$\frac{N_c(N_c-1)}{2}$ if $a= \pm 2\Lambda^{2N_c+2-N_f}_{N=2}$ 
if we recover the scale factor. 
The former coincides with the 
$(r=\frac{N_f-1}{2})$-th Higgs branch and the latter coincides with the 
dimension of the $(r=\frac{N_f}{2})$-th Higgs branch. 
Also this result agrees with the findings of \cite{Hori2};
the $(r=\frac{N_f}{2})$-th Higgs branch 
emanates from the locus where only 
$\frac{N_f-2}{2}$ of $\phi_a$'s vanish and the product of nonvanishing 
$\phi_a^2$'s is $\pm 2\Lambda^{2N_c+2-N_f}$. 
In \cite{Hori2}, this result is obtained using $O4$-plane while 
we use $O6$-plane. 
For $r=\frac{N_f-1}{2}$, similar calculation shows that the Higgs-branch 
emanates from the locus where all $\frac{N_f-1}{2}$ of $\phi_a$'s 
vanish and gives the correct dimension $\frac{N_f(N_f-1)}{2}$.  
  
\subsection{The N=2 Higgs branch of $SO(2N_c)$ gauge theory}
Our understanding of the Higgs branch of the N=2 $SO(2N_c)$ gauge theory 
is incomplete, since we do not have a good understanding of the M-theory 
geometry corresponding to $O6^+$-orientifold. 
According to \cite{Argyres}, the (mixed Coulomb)-Higgs branches are 
again classified by an integer $r=1,2,\cdots [\frac{N_f}{2}]$ where 
$N_f$ is the number of hypermultiplets. 
The $r$-th Higgs branch has quaternionic dimension $2rN_f-(2r^2-r)$ 
and emanates from a $(N_c-r)$-dimensional complex subspaces of the 
Coulomb branch, where there is an unbroken gauge group $SO(2r)$.
Since $SO(2N_c)$ gauge theory are symptotically free when 
$N_f \leq 2N_c-2$, the $SO(2r)$ factors at the root of the 
$r$-branches are all infrared-free, and will remain unbroken 
quantum mechanically.  
The lifting of D6-brane configuration 
is described by $xy=\Lambda^{4N_c-4-2N_f}v^{4+2N_f}$ in M-theory 
and the space has $D_{4+N_f}$ singularity where $D_4$ singularity 
cannot be blownup. This means that we can blowup the singularity 
until we get at $D_4$ singularity. Again the curve for the Coulomb branch 
is given by $y=\frac{1}{2}B_{N_c}(z)$. 
If we have ordinary $D_{4+N_f}$ singularity, 
then we can have $N_f+4$ blownup rational curves 
$C_1, C_2 \cdots, C_{N_f}, C_{N_f+1}, C_{N_f+2}, C_{N_f+3}, C_{N_f+4}$. 
If we have $F=y-z^r$ near singularity, we obtain the zeroes of 
$1,2, \cdots 2r \cdots 2r, 2r,2r,r,r$ from $C_1,C_2, \cdots, C_{2r}, 
\cdots C_{N_f}, C_{N_f+1}, C_{N_f+2}, C_{N_f+3}, C_{N_f+4}$ 
respectively. If the $D_4$ singularity cannot be blownup, we should 
exclude the zeroes obtained from $C_{N_f+1}$ to $C_{N_f+4}$.
This is especially obvious when we just consider the pure $SO(2N_c)$ 
gauge theory with $N_f=0$. 
Since this theory does not have a Higgs branch, 
we should not have any contribution from the frozen $D_4$ singularity. 
Thus the quaternionic dimension of the $r$-th Higgs branch is given by 
$\Sigma_{i=1}^{2r}i+2r(N_f-2r)=2rN_f-2r^2+r$ as desired. 

\begin{figure}
\centering
\epsfxsize=5in
\hspace*{0in}
\epsffile{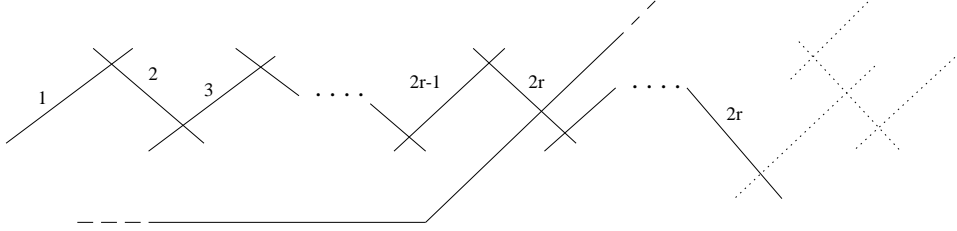}
\caption{The $r$-th Higgs branch of $SO(2N_c)$ gauge theory in M-theory. 
$D_4$ singularity which would be blown up in the usual $D_{4+N_f}$ 
singularity is represented by dotted lines.}
\label{fig:so}
\end{figure}

Unlike the case of $Sp(N_c)$ case where we can understand 
$SO(2N_f)$ flavor in various ways in M-theory, we do not understand 
how the $Sp(N_f)$ flavor symmetry is realized for $SO(2N_c)$ gauge 
group in M-theory picture. 
This requires better understanding of $D_{4+N_f}$ singularity with 
frozen $D_4$ singularity. One intriguing fact is that the geometry 
of usual $D_{4+N_f}$ singularity and that of $D_{4+N_f}$ singularity 
with frozen $D_4$ singularity are the same. The geometry is determined 
uniquely by the hyperk\"{a}hler property of space metric in M-theory
\cite{SeWi}. 
There should be some additional ingredient in M-theory which freezes 
$D_4$ singularity thereby producing Sp-symmetry, but currently 
that ingredient is not well understood. 

\medskip

\section{N=1 rotated curve}
\subsection{$Sp(N_c)$ gauge group}
We can break N=2 to N=1 by giving a bare mass $\mu$ to the adjoint
chiral multiplet. For small $\mu$, we can use the Seiberg-Witten 
curve to analyze the structure of vacua. Such analysis is done in 
\cite{Argyres}. Most of the Coulomb branch is lifted except for 
discrete points where the underlying curve is degenerated 
into a curve of genus zero.  
The remaining vacua are in the locus where $r=[\frac{N_f-1}{2}]$ 
of the parameters $\phi_a$ in 
$B_{N_c}=\prod_{a=1}^{N_f}(v^2-\phi_a^2)$
 vanish as well as in the locus where 
$r=N_f-N_c-2$ of them vanishing. The former is called the $A$-branch root 
and the latter is called the $B$-branch root in\cite{Hori2}. 
The $B$-branch is a singlet which is invariant under discrete 
$Z_{2N_c+2-N_f} R$-symmetry,
where as the $A$-branch root consists of $2N_c+2-N_f$ points. 
For $N_f$ odd these are related by $Z_{2N_c+2-N_f}$, but for 
$N_f$ even they fall into two separate orbits, each having 
$N_c+1-\frac{N_f}{2}$ points. On the other hand, pure Yang-Mills 
theory has $N_c+1$ supersymmetric vacua related by the discrete 
$R$-symmetry. 

In the Type IIA setup with O6-plane describing $SO/Sp$ gauge theories, 
we can rotate the left NS5-brane by $\theta$ and the right NS5-brane 
by $-\theta$. This corresponds to giving a mass for the adjoint chiral 
multiplet in N=2 theory, thereby reducing N=2 to N=1\cite{Barbon}. 
Both of the 5-branes are extended in the $x^8, x^9$-directions, 
and we need additional complex coordinates 
\begin{equation}
w=x_8+ix_9
\end{equation}
in order to describe the corresponding configuration of the M5-brane.
Since the two NS5-branes correspond to the two asymptotic 
regions with $v\rightarrow \infty$, where $y\sim v^{2N_c}$ 
and $x\sim v^{2N_c}$ respectively. 
Thus the rotation requires the boundary conditions
\begin{eqnarray}
w\rightarrow \mu v, & &  y\sim v^{2N_c}\sim\lambda^{2N_c}, \,\,\,\,\,\, 
v\rightarrow \infty  \\
w\rightarrow -\mu v,& &  x\sim v^{2N_c}, 
y\sim v^{-4+2N_f-2N_c}\sim \lambda^{4-2N_f+2N_c},  
v\rightarrow \infty.
\end{eqnarray}
We can identify $\mu$ as the mass of the adjoint chiral multiplet 
by using $R$ symmetry. The N=2 configuration is invariant under 
the rotation $U(1)_{4,5}$ and $SU(2)_{7,8,9}$. After the rotation, 
$SU(2)_{7,8,9}$ is broken to $U(1)_{8,9}$ if $\mu$ is assigned $(-2,2)$ 
charge of $U(1)_{4,5}\times U(1)_{8,9}$. 
Since this is the same as the $R$-charge of the mass of the adjoint 
field and since there is no other parameters charged under $U(1)_{8,9}$, 
the two quantities should be identified. 

Consider $Sp(N_c)$ gauge theory with $N_f$ hypermultiplets. 
The charges of the coordinates and parameters are the following. 
\begin{equation}
\begin{array}{ccc}
 &U(1)_{4,5} & U(1)_{8,9} \\
v & 2 & 0 \\
w & 0 & 2 \\
y & 4N_c & 0 \\
x & 4N_c & 0 \\
\mu & -2 & 2 \\
\Lambda^{2N_c+2-N_f}_{N=2} & 4N_c+4-2N_f & 0 
\end{array}
\end{equation}
The combination of $\frac{1}{2}U(1)_{4,5}+\frac{1}{2}U(1)_{8,9}$ 
makes $\mu$ invariant, and its $Z_{2N_c+2-N_f}$ subgroup makes 
$\Lambda^{2N_c+2-N_f}_{N=2}$ invariant. 
In field theory, the mass of the adjoint chiral multiplet lifts the 
Coulomb branch. Thus rotating the branes can only be done 
at points where the curve degenerates to a curve of genus zero. 
Then we can introduce an auxiliary complex parameter $\lambda$ 
and identify the genus zero curve as the complex $\lambda$ plane
with some points deleted or points at infinity added. 
$v,w$ and $t$ can be expressed as a rational function of 
$\lambda$\cite{Witten5}. 
Here we 
assume that the rotated curve consists of a single 
component. The case where the curve consists of more than one 
components will be treated separately.
For rotated curve $C$, we can consider $\tilde{C}$, the projection 
of $C$ into $x,y,v$ space. 
Since $\mu$ is the only parameter, we cannot deform the N=2 curve 
equation without breaking $U(1)_{8,9}$ symmetry. Thus the projection 
of the rotated curve $\tilde{C}$ on $x,y,v$ remains the same N=2 curve
\cite{Ooguri}. 
$v$ has two poles 
corresponding to the two NS5-branes and we take two poles to occur at 
$\lambda=0$ and $\lambda=\infty$. Thus we have 
\begin{equation}
v=b(\lambda+\frac{1}{\lambda}). \label{eq:v}
\end{equation}
There might be a constant term in the right hand side, but this should 
vanish if we take $Z_2$ acts as $\lambda\rightarrow -\frac{1}{\lambda}$.
We take the asymptotic conditions as 
\begin{eqnarray}
w\sim \mu v, & & \lambda\rightarrow \infty \\
w\sim -\mu v, & & \lambda\rightarrow 0. \nonumber
\end{eqnarray}
This determines 
\begin{equation}
w=\mu b(\lambda-\frac{1}{\lambda}).
\end{equation}
Thus $w$ is invariant under the $Z_2$ action $\lambda\rightarrow 
-\frac{1}{\lambda}$ while $v$ is odd. Asymptotic conditions for $x,y$ 
are given by 
\begin{eqnarray}
 y\sim v^{2N_c}\sim\lambda^{2N_c}, & & \,\,\,\,\,\, 
\lambda\rightarrow \infty  \\
x\sim v^{2N_c}, & & y\sim v^{-4+2N_f-2N_c}\sim \lambda^{4-2N_f+2N_c},  
\lambda\rightarrow 0
\end{eqnarray}
where $x,y$ satisfy (\ref{eq:D61}) and (\ref{eq:sp}). 
Thus $y=\lambda^{2N_c+4-2N_f} \frac{P_{2N_f}(\lambda)}{P_4(\lambda)}$
where $P_i(\lambda)$ is the $i$-th polynomial of $\lambda$. 
Now under the $Z_2$ action $\lambda\rightarrow -\frac{1}{\lambda}$, 
$y$ is maped to $x$ so
$x=(\frac{1}{\lambda})^{2N_c+4-2N_f} 
\frac{P_{2N_f}(-\frac{1}{\lambda})}{P_4(-\frac{1}{\lambda})}$.
The expression should satisfy (\ref{eq:D61}).  
\begin{eqnarray}
xy&=&\Lambda^{4N_c+4-2N_f}_{N=2} v^{-4}\prod_{i=1}^{N_f} (v^2-m_i^2) \\
 &=&  \Lambda^{4N_c+4-2N_f}_{N=2} b^{-4+2N_f} 
(\lambda+\frac{1}{\lambda})^{-4}\prod_{i=1}^{N_f} 
((\lambda+\frac{1}{\lambda})^2-\frac{m_i^2}{b^2}).
\end{eqnarray}
From this, we can set $P_4(\lambda)=\lambda^2
(\lambda+\frac{1}{\lambda})^2$ and 
$P_{2N_f}(-\frac{1}{\lambda})P_{2N_f}(\lambda)
=\Lambda^{4N_c+4-2N_f}_{N=2} b^{-4+2N_f} \prod_{i=1}^{N_f}
((\lambda+\frac{1}{\lambda})^2-\frac{m_i^2}{b^2})$. 
This can be satisfied if we choose
$P_{2N_f}(\lambda)=C \prod_{i=1}^{N_f}(\lambda^2-a_i^2)$ 
with $a_i+\frac{1}{a_i}=\pm \frac{m_i}{b}$ and 
$C^2(-1)^{N_f}\prod_{i=1}^{N_f}a_i^2= 
\Lambda^{4N-c+4-2N-f}_{N=2} b^{-4+2N_f}$. 
We choose 
\begin{equation}
a_i=\frac{2}{\frac{m_i}{b}+\sqrt{(\frac{m_i}{b})^2-2}}. \label{eq:a} 
\end{equation}

Therefore we have 
\begin{eqnarray}
y&=& \frac{\lambda^{2N_c+2-2N_f}C \prod_{i=1}^{N_f}(\lambda^2-a_i^2)}
{(\lambda+\frac{1}{\lambda})^2}  \label{eq:xy10} \\
x&=& \frac{(\frac{1}{\lambda})^{2N_c+2-2N_f}C \prod_{i=1}^{N_f}
((\frac{1}{\lambda})^2-a_i^2)}{(\lambda+\frac{1}{\lambda})^2}. \nonumber
\end{eqnarray}
From $x+y=B_{N_c}(v^2)=v^{2N_c}+\cdots$, we have $C=b^{2N_c}$ comparing 
the highest order term in $\lambda$. 
Therefore we obtain 
\begin{equation}
b^{4N_c-2N_f+4}=\frac{\Lambda^{4N_c+4-2N_f}_{N=2}} 
{(-1)^{N_f}\prod_{i=1}^{N_f}a_i^2}. \label{eq:b}
\end{equation}
Note that with $a_i$ in (\ref{eq:a}), in the limit of 
$m_{N_f}\rightarrow \infty$, the expression (\ref{eq:b})  
is correctly reduced to $N_f-1$ case with the replacement 
\begin{equation}
\Lambda^{4N_c+4-2N_f}_{N=2}(-m_i^2)\rightarrow 
\Lambda^{4N_c+4-2(N_f-1)}_{N=2}.
\end{equation}
This is consistent with the one-loop matching of the gauge coupling.

In case $m_i=0$, $a_i=\pm i$ and 
$b^{4N_c+4-2N-f}_{N=2}= \Lambda^{4N_c+4-2N-f}_{N=2}$. 
One can check that with the expression (\ref{eq:xy10}), there's a 
special $u_i$'s  for $B_{N_c}(v^2)=v^{2N_c}+u_1v^{2N_c-2}+ \ldots 
u_{N_c}$ where 
$x+y=B_{N_c}(v^2)+\frac{\prod_{j=1}^{N_f}im_j}{v^2}$ holds.
$b^2$ has $2N_c+2-N_f$ solutions. 
The $Z_{2N_c+2-N_f}$ rotational symmetry is completely broken 
by $b^2$ for odd $N_f$ and is broken to $Z_2$ for even $N_f$, 
since its generator acts $b^2$ as 
$b^2\rightarrow e^{\frac{4\pi i}{2N_c+2-N_f}}b^2$. 
For $N_f=0$, apparently we obtain $2N_c+2$ solutions from (\ref{eq:b}). 
But we know that there are only $N_c+1$ vacua for pure Yang-Mills 
theory. The apparent discrepancy is resolved if we carefully look at 
the equation $x+y=B_{N_c}(v^2)+\frac{2\Lambda^{2N_c+2}_{N=2}}{v^2}$. 
$b$ should satisfy $b^{2N_c+2}=(-1)^{N_c}\Lambda^{2N_c+2}_{N=2}$ 
in order to cancel the $\frac{1}{v^2}$ term. 

The above result is consistent with the N=1 vacua of the $Sp(N_c)$ 
gauge theory coming from the $A$-branch. The structure of the above 
solution is similar to \cite{Hori2}, and we can follow their argument. 
One can show that $B_{N_c}(v^2)$ defining the projected curve in 
$x$-$y$-$v$ plane is of the form 
$B_{N_c}(v^2)=cv^{2r}+\cdots , c\neq 0,$ where 
$r\geq \frac{N_f-1}{2}$ for $N_f$ odd and $r=\frac{N_f-2}{2}$ for 
$N_f$ even which is the property of the $A$-branch. 
This can be seen as follows. As $\lambda \rightarrow \pm i$, 
$v\rightarrow 0$ and $y\rightarrow v^{N_f-2}$ 
as can be seen from (\ref{eq:v}) and (\ref{eq:xy10}). 
On the other hand, equation (\ref{eq:sp}) implies 
\begin{equation}
y\sim -\frac{cv^{2r}}{2}\pm\sqrt{\frac{c^2v^{4r}}{4}
-\Lambda^{4N_c+4-2N_f}v^{2N_f-4}}.  \label{eq:rbranch}
\end{equation}  
The two statements are consistent only when $r\geq\frac{N_f-2}{2}$. 
In particular, if $N_f$ is even, 
equation (\ref{eq:v}) and (\ref{eq:xy10}) shows that 
$y$ is a single valued function of $v$ near $v\sim 0$. 
This is possible only if the two terms in the square root of 
(\ref{eq:rbranch}) cancel. Thus $r=\frac{N_f-2}{2}$ and 
$c=\pm 2\Lambda^{2N_c+2-N_f}_{N=2}$. 

So far we assume that the infinite curve consists of a single component. 
But suppose the curve is factorized so that the infinities are separated 
in covering space. When such factorization occurs, we just have 
to rotate each component. 
One can see that the factorization is unique and is given by
\begin{equation}
(y-v^{2N_c})(y-\lambda^{4N_c+4-2N_f}_{N=2}v^{2N_f-2N_c-4})=0. 
\end{equation}
This gives $B_{N_c}(v^2)=v^{2N_c}+\Lambda^{4N_c+4-2N_f}$. 
From this, we see that the factorization is possible only when 
$N_f\geq N_c+2$. Also this shows that the curve belongs to 
$r=N_f-N_c-2$ Higgs branch. The rotated curve is just the union of 
\begin{eqnarray}
y&=&v^{2N_c} \\
x&=&\Lambda^{4N_c+4-2N_f}v^{2N_f-2N_c-4} \nonumber \\
w&=&\mu v \nonumber
\end{eqnarray}
and 
\begin{eqnarray}
x&=&v^{2N_c} \\
y&=&\Lambda^{4N_c+4-2N_f}v^{2N_f-2N_c-4} \nonumber \\
w&=&-\mu v \nonumber
\end{eqnarray}
\subsection{$SO(2N_c)$ gauge theory}
Similar analysis can be done for $SO(2N_c)$ case. 
First consider the rotated curve consisting of one component in covering 
space. 
For such a rotated curve $C$, the projection of the rotated curve 
$\tilde{C}$ on $x,y,v$ remains the same N=2 curve. 
$v$ and $w$ are given by the same expressions as before
\begin{eqnarray}
v&=&b(\lambda+\frac{1}{\lambda}). \label{eq:vw} \\
w&=&\mu b(\lambda-\frac{1}{\lambda}). \nonumber
\end{eqnarray}
 Asymptotic conditions for $x,y$
are given by
\begin{eqnarray}
y\sim v^{2N_c}\sim\lambda^{2N_c}, & & \,\,\,\,\,\, 
\lambda\rightarrow \infty  \\
x\sim v^{2N_c},& &  y\sim v^{4+2N_f-2N_c}\sim \lambda^{-4-2N_f+2N_c}, 
\lambda\rightarrow 0 \nonumber
\end{eqnarray}
where $x,y$ satisfy (\ref{eq:xyso}) and (\ref{eq:so}). 
By the similar calculation to $Sp(N_c)$ case, we obtain 
\begin{eqnarray}
y&=& \lambda^{2N_c-2-2N_f}(\lambda+\frac{1}{\lambda})^2 C 
\prod_{i=1}^{N_f}(\lambda^2-a_i^2)
 \\ \label{eq:xy2}
x&=& (\frac{1}{\lambda})^{2N_c-2-2N_f}
(\lambda+\frac{1}{\lambda})^2 C \prod_{i=1}^{N_f}
((\frac{1}{\lambda})^2-a_i^2) \nonumber
\end{eqnarray}
with $C^2(-1)^{N_f}\prod_{i=1}^{N_f}a_i^2=
\Lambda^{4N_c-4-2N_f}_{N=2} b^{4+2N_f}$ and 
\begin{equation}
b^{4N_c-2N_f-4}=\frac{\Lambda^{4N_c-4-2N_f}_{N=2}}
{(-1)^{N_f}\prod_{i=1}^{N_f}a_i^2}. \label{eq:b3} 
\end{equation}
In the limit of $m_{N_f}\rightarrow \infty$, we correctly reproduce 
the formula for $N_f-1$ case. 
If all $m_i$ vanish in (\ref{eq:xyso}), we have 
$b^{4N_c-2N_f-4}=\Lambda^{4N_c-4-2N_f}_{N=2}$ and $b^2$ has 
$2N_c-2-N_f$ solutions. The $Z_{2N_c-2-N_f}$ rotational symmetry is 
completely broken by $b^2$ for odd $N_f$ and is broken to 
$Z_2$ for even $N_f$. 
For $N_f=0$, we obtain $2N_c-2$ solutions. This agrees with the Witten 
index calculation for $N_c\geq 3$, For $N_c=2$, we have 4 vacua since  
$SO(4)\sim SU(2) \times SU(2)$ and each $SU(2)$ gives two vacua. 
Thus we should have two more vacua, which arise in the 
reducible component as we will see shortly.

By the similar argument to $Sp(N_c)$ gauge group, we can see that 
$B_{N_c}(v^2)$ defining the projected curve in $(x,y,v)$ plane is of the 
form $B_{N_c}(v^2)=cv^{2r}+\cdots, c\neq 0$ where $r\geq 
\frac{N_f+2}{2}$. 
For $N_f$ even, $r=\frac{N_f+2}{2}$ and  
$c=\pm 2\Lambda_{N=2}^{2N_c-2-N_f}$. 
For $N_f$ even, the maximal value $r$ allowable for the 
Higgs branch is $\frac{N_f}{2}$. Thus in order to have 
$B_{N_c}(v^2)=v^{N_f}v^2(v^{2N_c-N_f-2}+\cdots)$ the $v^2$ factor 
should be attributed to the Coulomb branch. This implies 
that the N=1 vacua are emanated from the $\frac{N_f}{2}$-th Higgs branch 
where the residual Coulomb branch is constrained to the 
$N_c-1-\frac{N_f}{2}$-dimensional subspace.

If the infinite curve is reducible in covering space, 
it is uniquely given by
\begin{equation}
(y-v^{2N_c})(y-\Lambda^{4N_c-4-2N_f}_{N=2}v^{2N_f-2N_c+4})=0.
\end{equation}
and the rotated curve is given by the union of  
\begin{eqnarray}
y&=&v^{2N_c} \label{eq:red1}\\
x&=&\Lambda^{4N_c-4-2N_f}v^{2N_f-2N_c+4} \nonumber \\
w&=&\mu v \nonumber
\end{eqnarray}
and
\begin{eqnarray}
x&=&v^{2N_c} \label{eq:red2}\\
y&=&\Lambda^{4N_c-4-2N_f}v^{2N_f-2N_c+4} \nonumber \\
w&=&-\mu v. \nonumber
\end{eqnarray}
Alternative we can have $w=-\mu v$ in (\ref{eq:red1}) and 
$w=\mu v$ in (\ref{eq:red2}). These two possibilities are related by 
discrete $R$ symmetry. This reducible curve exists for $N_c=2$ even 
with $N_f=0$ and two possibilities mentioned above provide two 
additional vacua for pure $SO(4)$ gauge theory with N=1 supersymmetry. 

In \cite{Csaki}, there appear the expressions for rotated curves of 
$SO(2N_c)/Sp(N_c)$ gauge theories. Since their Type IIA configuration 
is the same as our configuration, 
their results should be consistent with ours. 
Indeed, from (\ref{eq:vw}), we have $(w+\mu v)(w-\mu v)=-4\mu^2b^2$ 
if we eliminate $\lambda$. This agrees with (5.3) of \cite{Csaki}. 
Also if  we change the variables, 
\begin{eqnarray}
\tilde{y}=\frac{y}{\prod_{i=1}^{N_f}(v+m_i)}
&=&\lambda^{2N_c-2}v^2
\frac{C\prod_{i=1}^{N_f}(1-\frac{a_i}{\lambda})}
{b^{N_f+2}\prod_{i=1}^{N_f}(\lambda+\frac{1}{a_i})} \\
 &=& v^2\lambda^{2N_c-2}
\frac{\Lambda^{2N_c-2-N_f}\prod_{i=1}^{N_f}(1-\frac{a_i}{\lambda})}
{(-1)^{\frac{N_f}{2}}\prod_{i=1}^{N_f}a_i
\prod_{i=1}^{N_f}(\lambda+\frac{1}{a_i})}
\end{eqnarray}
using (\ref{eq:b3}) (\ref{eq:vw}), 
the expression coincides with (5.4) of 
\cite{Csaki} up to constant which can be absorbed into 
$\Lambda_{N=2}$ in our expression. And 
$\tilde{x}=\frac{x}{\prod_{i=1}^{N-f}(v-m_i)}$ gives the corresponding 
expression in \cite{Csaki}. 
Note that $\tilde{x}\tilde{y}\propto v^4$. 
This change of variable corresponds to moving D6-branes outside of 
NS5-branes, thereby creating semi-infinite D4-branes. 
Indeed from (\ref{eq:so2}), we have after the change of variables, 
\begin{equation}
\tilde{y}^2\prod_{i=1}^{N_f}(v+m_i)-B_{N_c}(v^2)\tilde{y}
+v^4\prod_{i=1}^{N_f}(v-m_i)=0.
\end{equation}
which describes $SO(2N_c)$ gauge theory where hypermultiplets are 
realized by semi-infinite D4-branes in Type IIA setting. 
Thus our expression represents M-theory lifting of the Type IIA 
configuration with D6-branes while the expressions in \cite{Csaki}
represents M-theory picture where D6-branes are moved to infinity, 
thereby producing semi-infinite D4-branes. 
In \cite{Csaki}, there are several consistent checks for their 
expressions. They checked that in a suitable limit, the expressions 
recover the Type IIA brane configuration they started with. They checked 
that decoupling of a flavor works correctly and the expressions 
encode N=1 duality properly. All those checks in \cite{Csaki} 
can be consistent checks 
for our expressions as well. 
The same relation holds for the rotated curve for $Sp(N_c)$ gauge 
group.
\medskip

\section{Chiral gauge theory}

\begin{figure}
\centering
\epsfxsize=5in
\hspace*{0in}
\epsffile{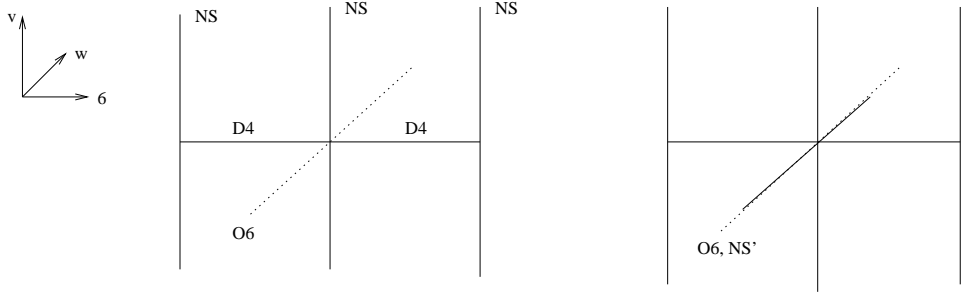}
\caption{The left figure is the brane configuration 
of $SU(2N_c)$ gauge theory with 
symmetric matter. The right figure is the chiral gauge theory 
configuration.}
\label{fig:brane4}
\end{figure}

In order to get the M5-brane picture of the chiral gauge theory, 
it's better to start with the N=2 $SU(2N_c)$ gauge theory with 
symmetric matter and consider the situation where the middle NS5-brane 
decouples. The corresponding brane configuration is depicted in 
Fig \ref{fig:brane4}. This theory is explored in detail in \cite{LL}. 
The bifundamentals of the product gauge group gives the symmetric 
matter under the orientifold projection.  
The curve describing Coulomb branch is given by
\begin{equation}
y^3+y^2\prod_{i=1}^{2N_c}(v-a_i)+yv^2\prod_{i=1}^{2N_c}(v+a_i)+v^6=0
\end{equation}
with $xy=v^4$.\footnote{Again, we set $\Lambda_{N=2}=1$.}
The coefficient in front of $y^2$ describes the position of 
D4-branes on the left of 
the middle NS5-brane and the coefficient in front of $y$ describes 
the D4-branes on the right of the 
middle NS5-brane in the classical limit. 
Generically the positions of left D4-branes in $v$ are different from the 
positions of right D4-branes and we cannot pull away the middle 
NS5-brane in $x^7$-direction without breaking gauge symmetry. 

However, if the curve has the special form
\begin{equation}  
y^3+y^2\prod_{i=1}^{N_c}(v^2-a_i^2)
+yv^2\prod_{i=1}^{N_c}(v^2-a_i^2)+v^6=0,
\end{equation}
the positions of left D4-branes match the positions of right D4-branes.
In this case, a left D4-brane and the corresponding right D4-brane can be 
recombined into a single 4-brane and the middle 5-brane decouples. 
This means that the curve is factorized as 
\begin{equation}
(y+v^2)(y^2+(\prod_{i=1}^{N_c}(v^2-a_i^2)-v^2)y+v^4)=0.
\end{equation}
Without the orientifold projection, this describes the special case of 
$SU(2N_c)\times SU(2N_c)$ gauge theory. The first factor corresponds 
to the middle NS5-brane and we can rotated it as we wish. 
But due to the orientifold projection, we have only two possibilities 
for the middle 5-brane. One is the NS5-brane and the other is 
NS$^{\prime}$5-brane where the rotated angle is $\frac{\pi}{2}$. 
And in both cases, central 
NS5-brane and NS$^{\prime}$5-brane are described by 
a decoupled rational curve. If the middle 5-brane is a 
NS$^{\prime}$5-brane, the brane configuration describes the chiral 
gauge theory. 

If we look at the brane configuration of the chiral gauge theory, 
one can see that each left D4-branes match a right D4-branes 
and recombine 
into a single D4-brane. 
Again NS$^{\prime}$5-brane is decoupled and is described by a rational 
curve in M-theory picture. 
Thus the suggested curve for the chiral gauge theory is the union 
of $v=0,x=0, y=0$ (isomorphic to $w$-plane) 
 and 
\begin{equation}
y^2+(\prod_{i=1}^{N_c}(v^2-a_i^2)-v^2)+v^4=0. \label{eq:c1}
\end{equation}
with $w=0$.
An intersting thing is that the second factor can be interpreted as two 
ways; N=2 $SO(2N_c)$ pure gauge theory or N=2 $Sp(N_c)$ gauge theory with 
4 fundamental N=2 massless hypermultiplets which are decomposed 
as 8 N=1 chiral 
multiplets.\footnote{Similar observation is made in \cite{Kuta}.} 
Recall that the general form of $Sp(N_c)$ curve is 
\begin{equation}
y^2+(B_{N_c}(v^2)+2A)y+v^{-4}\prod_i^{N_f}(v^2-b_i^2)=0 \label{eq:c2}
\end{equation}
with $A=\frac{2\prod_i^{N_f}b_i}{v^2}$. 
If we have a massless hypermultiplets 
then some $b_i$ vanishes to give $A=0$.   

One can easily incorporate the D6-branes.
The curve is a union of a rational curve isomorphic to $w$-plane and 
\begin{equation}
y^2+B_{N_c}(v^2)y+v^4\prod_{i=1}^{N_f}(v^2-a_i^2)=0 \label{eq:c3}
\end{equation}
with $w=0$. Again (\ref{eq:c3}) can be interpreted as 
N=2 $SO(2N_c)$ gauge theory 
with $N_f$ flavors or $Sp(N_c)$ gauge theory with $N_f+4$ hypermultiplets 
where 4 of them are massless. This is consistent with 
the Type IIA brane configuration. If the middle NS$^{\prime}$5-brane 
is located at 
$x^7=0$, we have the chiral gauge theory. But if the NS$^{\prime}$5-brane 
is moved to negative direction, we have the N=2 $Sp(N_c)$ gauge theory 
with $N_f+4$ hypermultiplets. If the NS$^{\prime}$5-brane 
is moved to positive direction, we have the N=2 $SO(2N_c)$ gauge 
theory with $N_f$ hypermultiplets. Also we have seen how the N=2 Higgs 
branch of the SO/Sp gauge theory could be understood in M-theory 
picture. These Higgs branches are a part of the moduli space of 
the chiral gauge theory. 

Since the NS$^{\prime}$5-brane is decoupled, the rotated curve for the 
chiral gauge theory can be obtained by attaching a rational curve 
to the rotated curve of (\ref{eq:c3}). Again the rotated curve for 
$SO(2N_c)$ gauge theory with $N_f$ flavors is the same as that for 
$Sp(N_c)$ gauge theory with $N_f+4$ flavors where four of them are 
massless hypermultiplets.

In \cite{Ha3}, using the brane construction of the chiral gauge theory 
the chiral non-chiral transition is discussed. We can put two additional 
NS$^{\prime}$5-branes on the O6-plane. Those NS$^{\prime}$5-branes are 
constrained to move in $x^7$-direction. Only when a pair of 
NS$^{\prime}$5-branes coincide, they can move in $x^4,x^5,x^6$-directions
pairwise. Once such movement occurs, we have non-chiral gauge theory 
with gauge group either $SU(2N_c)\times Sp(N_c)$ or 
$SU(2N_c)\times SO(2N_c)$ depending on the sign of the orientifold 
D4-branes intersect. The matter contents of the resulting 
theory are a symmetric/antisymmetric tensor for the $SO/Sp$ group and 
a pair (chiral and its conjugate) of bifundamental fields. 
From the discussions above, we can describe the chiral gauge theory 
configuration where two additional NS$^{\prime}$-branes are located 
on orientifold. Since NS$^{\prime}$5-branes are factorized 
in the chiral gauge theory setup, we can just
attach two rational curves in $x^7$ direction to the curve (\ref{eq:c3}). 
Again as we traverse a NS$^{\prime}$5-brane, the interpretation 
of the ambient space $xy=v^4$ is changed. In one case, we have 
Atiyah-Hitchin space with the lifting of 8 half D6-branes. 
In the other case, we have frozen $D_4$ singularity. 
We can make two NS$^{\prime}$5-branes coincident, thereby producing 
a nontrivial fixed point which mediates chiral non-chiral transition. 
But since we only know the low energy behavior of the nontrivial 
fixed point, not much can be said about the fixed point just using 
M-theory description obtained in this paper. 
It is also quite interesting to understand 
the other non-chiral branch in M5-brane picture. This involves 
the understanding of the rotation of NS5-branes in N=2 configuration by 
$\frac{\pi}{2}$ in M-theory picture.

\begin{figure}
\centering
\epsfxsize=5in
\hspace*{0in}
\epsffile{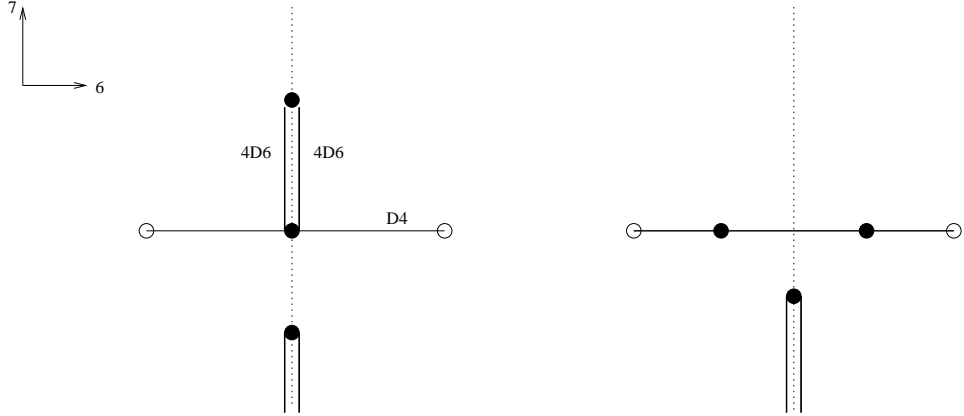}
\caption{Chiral nonchiral transition from the $SU(2N_c)$ chiral theory 
to the $SU(2N_c)\times SO(2N_c)$ nonchiral theory. Black circles 
denote NS$^{\prime}$5-branes and white circles denote NS5-branes.}
\label{fig:transition}
\end{figure}

\medskip

\section{Discussion}

We have explored the M5-brane realization of one specific chiral 
gauge theory. Since the Type IIA brane configuration of the chiral 
gauge theory is constructed using orientifold 6-plane, a part of 
the paper is devoted to studying the corresponding M-theory 
configuration of the orientifold 6-planes. 
The geometric interpretation in M-theory of the orientifold 
with $-4$ RR charge ($O6^-$) is the Atiyah-Hitchin space. This 
clear geometric interpretation makes easy 
the analysis of the gauge theory 
constructed using $O6^-$-plane.  
On the other hand, the corresponding geometry in M-theory of the 
orientifold of +4 RR charge ($O6^+$) is less clear and we do not have 
 the complete understanding of the gauge theory arising from it. 
The $O6^+$-plane is lifted to the $D_4$ singularity which cannot 
be blownup, but we do not understand why it cannot be blownup. 
Related problem is that we cannot distinguish geometrically 
usual $D_4$-singularity which corresponds to 4 D6-branes on top of 
the $O6^-$-plane and the frozen $D_4$ singularity corresponding 
to the $O6^+$-plane. Similar problem arise if we consider 
the $SO(2N_c)/Sp(N_c)$ gauge theory with $N_f/(N_f+4)$ with 4 
massless fundamental flavors. The same curve describing the both 
theories and by moving the detached rational curve from the chiral 
gauge theory configuration, we can move to either theories. 
But we cannot tell in which case we have the degrees of freedom 
corresponding to the physical D6-branes just by looking the curve. 
To resolve this issue would be in important step 
to understand M-theory beyond its 
supergravity approximation.

One special feature of the chiral gauge theory is the presence of half
D6-branes. It is argued in \cite{Hanany6} that the half D-branes 
play an important role in the realization of the chiral symmetry. 
But in usual brane configurations, half D-branes have the other 
half D6-branes to form whole D-branes. The configuration considered 
here is a rare case where half D6-branes do not have their pairs. 
And indeed these half D6-branes give the chiral spectrum in the 
gauge theory. It would be interesting to consider more examples where 
half D-branes which do not have their pairs 
give rise to a chiral spectrum.

{\bf Acknowlegment}

 We thank A. Hanany, M. Strassler, A. Uranga 
and E. Witten for useful discussions. We thank H. Ooguri for 
correspondence.


\newpage


\begin{thebibliography}{99}

\bibitem{HW} A. Hanany and E. Witten,
``Type IIB Superstrings, BPS Monopoles, And Three-Dimensional Gauge
Dynamics,''
hep-th/9611230, Nucl.Phys. B492 (1997) 152.

\bibitem{Witten1} E. Witten, ``Solutions of Four-Dimensional 
Field Theories
via $M$ Theory,'' hep-th/9703166, Nucl.Phys. B500 (1997) 3.

\bibitem{Witten3} E. Witten, ``Toroidal Compactification 
without Vector Structure,'' hep-th/9712028. 

\bibitem{Gi} A. Giveon, D. Kutasov, ``Brane Dynamics and Gauge Theory,''
hep-th/9802067 and references there in.

\bibitem{Johnson} N. Evans, C.V. Johnson and A.D. Shapere, 
``Orientifolds,
  Branes, and Duality of 4D Gauge Theories,'' hep-th/9703210,
  Nucl.Phys. B505 (1997) 251.

\bibitem{Hanany2} A. Hanany and A. Zaffaroni, ``On the Realization 
of Chiral Four-Dimensional Gauge Theories Using Branes,'' 
hep-th/9801134.

\bibitem{LL} K. Landsteiner and E. Lopez,
``New Curves from Branes,''
hep-th/9708118.

\bibitem{Trivedi} J. Lykken, E. Poppitz and S.P. Trivedi,
``Chiral Gauge Theories from D-Branes,''
hep-th/9708134; ``M(ore) on Chiral Gauge Theories from D-Branes,''
hep-th/9712193

\bibitem{Hanany4} A. Hanany, M. Strassler and A. Uranga, ``Finite 
Theories and Marginal Operators on the Brane,'' hep-th/9803086. 

\bibitem{LLL} K. Landsteiner, E. Lopez and D.A. Lowe, ``Duality
 of Chiral $N=1$ Supersymmetric Gauge Theories via Branes,''
 hep-th/9801002.

\bibitem{Ha3} I. Brunner, A. Hanany, A. Karch and D. L\"ust,
 ``Brane Dynamics and Chiral Nonchiral Transitions,''
 hep-th/9801017.

\bibitem{Kuta} S. Elitzur, A. Giveon, D. Kutasov and D. Tsabar,
  ``Branes, Orientifolds and Chiral Gauge Theories'',
  hep-th/9801020.


\bibitem{Ahn2} C. Ahn, K. Oh and R. Tatar, 
`` Comments on SO/Sp Gauge Theories from 
Brane Configuration with an O(6) Plane,'' hep-th/9803197.


\bibitem{LLL2} K. Landsteiner, E. Lopez and D. A. Lowe,
 private communication

\bibitem{Argyres} P.C. Argyres, M.R. Plesser and A.D. Shapere,
``N=2 Moduli Spaces and N=1 Dualities for $SO(n_c)$ and $USp(2n_c)$
SuperQCD,'' hep-th/9608129, Nucl.Phys. B483 (1997) 172-186.

\bibitem{Hori} K. Hori, H. Ooguri and C. Vafa,
``Non-Abelian Conifold Transitions and N=4 Dualities in Three
Dimensions,''
hep-th/9705220, Nucl.Phys. B504 (1997) 147-174.

\bibitem{SeWi} N. Seiberg and E. Witten,``Gauge Dynamics and 
Compactification to Three Dimensions,'' hep-th/9607163. 

\bibitem{Se} N. Seiberg,``IR Dynamics on Branes and Space-Time 
Geometry,'' hep-th/9606017, Phy.Lett. B384 (1996) 81-85.

\bibitem{AH} M. F. Atiyah and N. Hitchin, {\it The Geometry and 
Dynamics of Magnetic Monopoles} (Princeton University Press, 1988). 

\bibitem{Sen} A. Sen, ``A Note on Enhanced Gauge Symmetries in M and 
String Theory,'' hep-th/9707123. 

\bibitem{Hori2}J. de Boer, K. Hori, H. Ooguri and Y. Oz,
`` Branes and Dynamical Symmetry Breaking,'' hep-th/9801060. 

\bibitem{Barbon} J.L.F. Barbon, ``Rotated Branes and N=1 Duality,``
hep-th/9703051, Phys. Lett. B402 (1997) 59-63. 

\bibitem{Ooguri} K. Hori, H. Ooguri and Y. Oz,``Strong Coupling 
Dynamics of Four-Dimensional N=1 Gauge Theories from M Theory 
Five-Brane,'' hep-th/9706082, Adv. Theor. Math. Phys. 1 (1998)1-52.  

\bibitem{Witten5} E. Witten,``Branes and the Dynamics of QCD,''
hep-th/9706109, Nucl. Phys. B507 (1997) 658-690. 

\bibitem{Csaki} C. Csaki, M. Schmaltz, W. Skiba and J. Terning, 
``Gauge Theories with Tensors from Branes and Orientifolds,'' 
hep-th/9801207. 

\bibitem{Hanany6}J. Brodie and A. Hanany, ``Type IIA Superstrings, 
Chiral Symmetry and N=1 4-D Gauge Theory Dualities,'' hep-th/9704043, 
Nucl. Phys. B506 (1997) 157-182; 
Hanany and A. Zaffaroni, ``Chiral Symmetry from Type IIA Branes,'' 
hep-th/9706047, Nucl. Phys. B509 (1997) 145-168. 
\end{thebibliography}
\end{document}